# CHAIN EXTENSION OF A CONFINED POLYMER IN STEADY SHEAR FLOW

**Pinaki Bhattacharyya and Binny J. Cherayil,[#]**

Dept. of Inorganic and Physical Chemistry, Indian Institute of Science,

Bangalore-560012, INDIA

## ABSTRACT

The growing importance of microfluidic and nanofluidic devices to the study of biological processes has highlighted the need to better understand how confinement affects the behavior of polymers in flow. In this paper we explore one aspect of this question by calculating the steady-state extension of a long polymer chain in a narrow capillary tube in the presence of simple shear. The calculation is carried out within the framework of the Rouse-Zimm approach to chain dynamics, using a variant of a nonlinear elastic model to enforce finite extensibility. Under the assumption that the sole effect of the confining surface is to modify the pre-averaged hydrodynamic interaction, we find that the calculated fractional chain extension $x$ is considerably smaller than its value in the bulk. Furthermore, the variation of $x$ with a dimensionless shear rate (the Weissenberg number, Wi) is in good qualitative agreement with data from experiments on the flow-induced stretching of $\lambda$- phage DNA near a non-adsorbing glass surface.

[#] Corresponding author. Email: cherayil@ipc.iisc.ernet.in

## I. INTRODUCTION

As microfluidic devices find ever wider applications in the experimental study of complex biological and condensed phase phenomena,[1] it is becoming increasingly important to understand the combined effects of flow and confinement on single polymer dynamics.[2] Flow effects on polymers in the bulk have already been extensively explored experimentally and theoretically,[3-11] and many microscopic details of the conformational changes caused by flow fields have now been uncovered. Unfolding and elongation are the common response of initially compact equilibrium structures to flow, but depending on the precise nature of the applied field, this response can be accompanied by tumbling, folding, curling, defect formation and other unusual effects[4]. The nature of the flow also determines the extent of elongation: pure extensional flows, for instance, can often produce nearly full extension of the polymer at high enough flow rates, while shear flows can at most produce 40-50% of full extension, even at very high shear rates.[4]

The situation tends to become more complicated when fluid flow occurs near impenetrable surfaces,[12-21] and there seems to be much less quantitative experimental information about chain conformations under these conditions. A report by Fang et al.[20] suggests that in the presence of steady shear, the fractional extension of flexible $\lambda$-phage DNA molecules of roughly 20 $\mu$m length amounts to less than about 15% of its equilibrium confined value if the distance separating the polymer from the wall is on the order of a third or less of its total contour length (a finding in broad agreement with earlier results from the same group on the extension of phage DNA molecules under torsional shear flow near a glass surface.[19]) At these polymer-surface distances, there seems to be little or no dependence of the extension[20] (or, in the case of the simulations of Woo et al.,[18] the effective viscosity) on the imposed shear rate, and the concentration of chains there is reduced as well. The depletion of polymer concentration near a

surface – also documented in several other simulations,[16] – can be explained in terms of a mechanism in which the confining surface, when close enough to the polymer, "cuts off" a part of the long-ranged, solvent-mediated, segment-segment hydrodynamic interaction (the Oseen tensor), leading to an asymmetry in the flow patterns around individual monomers, and pushing them away from the surface towards the bulk.[20]

The reduction in the fractional chain extension near the surface also appears to originate in the hydrodynamics of the polymer-surface interaction. This interaction, by virtue of the no-slip boundary condition at the surface, tends to be screened on length scales on the order of the characteristic dimensions of the chain in free solution.[14,16] Under strong confinement, therefore, the polymer becomes effectively free-draining, i.e., Rouse-like, and in the absence of flow undergoes significant stretching in the direction parallel to the surface. Very little *additional* stretching therefore takes place if a flow field is now applied. Some evidence for this qualitative explanation of the reduction in the fractional chain extension near a surface can be found in experiments[14] and simulations,[15,16,18] but a well-defined analytical model of the underlying physics still seems to be lacking. Models based on an Edwards-type formalism for the coupled dynamics of polymer and solvent velocity field have already been developed for *unbounded* polymers,[22,23,24] and have been used successfully to interpret their response to shear, extensional and linear mixed flows.[7,8,25] The development of methods to incorporate hydrodynamic interactions into this formalism,[9,26] and to treat them in the presence of confining cylindrical surface[13] has now opened up the possibility of extending the formalism further, and applying it to the study of *surface*-mediated polymer-flow problems. This possibility is what we explore in the present paper by using the formalism to calculate the fractional extension of a continuum Gaussian polymer in a narrow capillary when subject to steady shear. The calculation shows that for this system the chain extension is indeed considerably smaller than its value in the bulk, and that its dependence on the shear rate (as expressed in terms of the Weissenberg number, Wi) is weak, in good qualitative agreement with data from experiments on the flow-induced stretching of $\lambda$- phage DNA near a non-adsorbing glass surface.[20]

Details of the calculation and its results are presented in the next several sections. Section II reviews the Rouse and Rouse-Zimm equations of chain dynamics, as well as their method of solution in terms of normal modes when a flow field is included and the chains are constrained to be finitely extensible. The constraint of finite extensibility is introduced via an ansatz motivated by so-called FENE models[23] of chain conformations. Section III applies the equations of Sec. II to the calculation of the chain dimensions of a polymer under shear in a narrow capillary, by evaluating the pre-averaged hydrodynamic interaction using an eigenfunction expansion, along the lines described by Harden and Doi.[13] Section IV discusses the results of these calculations, and presents some general conclusions.

## II. CHAIN DYNAMICS: THEORETICAL BACKGROUND

### A. Rouse and Rouse-Zimm approximations

In the present model, the conformation of a polymer of length $N$ is represented by a set of points $\{\mathbf{r}(\tau,t)\}$; each point $\mathbf{r}(\tau,t)$ specifies the spatial coordinates at time $t$ of a monomer on the chain backbone that is a distance $\tau$ from one end. The Hamiltonian $H_0$ of such a conformation, under theta solvent conditions, is given by[22,27]

$$H_0 = \frac{k_0}{2} \int_0^N d\tau \left( \frac{\partial \mathbf{r}(\tau,t)}{\partial \tau} \right)^2 \tag{1}$$

where $k_0$, the "spring constant" of the bond between adjacent monomers, is defined as $k_0 = 3k_B T / l_0^2$, $k_B T$ being the Boltzmann factor and $l_0$ the Kuhn length of the bond. These conformations evolve according to the well-known Rouse equation[22]

$$\zeta \frac{\partial \mathbf{r}(\tau,t)}{\partial t} = k_0 \frac{\partial^2 \mathbf{r}(\tau,t)}{\partial \tau^2} + \boldsymbol{\theta}(\tau,t) \qquad (2)$$

where $\zeta$ is the monomer friction coefficient, and $\boldsymbol{\theta}(\tau,t)$ is a random force (representing solvent fluctuations) that is defined by the following averages: $\langle \boldsymbol{\theta}(\tau,t) \rangle = \mathbf{0}$ and $\langle \theta_\alpha(\tau,t) \theta_\beta(\tau',t') \rangle = 2\zeta k_B T \delta_{\alpha\beta} \delta(\tau - \tau') \delta(t - t')$. The linear transformation $\mathbf{X}_p(t) \equiv N^{-1} \int_0^N d\tau \cos(p\pi\tau/N) \mathbf{r}(\tau,t)$ can be used to convert Eq. (2) to

$$\frac{\partial \mathbf{X}_p(t)}{\partial t} = -\frac{k_p}{\zeta_p} \mathbf{X}_p(t) + \frac{1}{\zeta_p} \mathbf{f}_p(t) \qquad (3)$$

where for $p \geq 1$, $\zeta_p = 2N\zeta$, $k_p = 2k_0 p^2 \pi^2 / N = 6k_B T p^2 \pi^2 / N l_0^2$ and $\mathbf{f}_p(t) = 2\int_0^N d\tau \cos(p\pi\tau/N) \boldsymbol{\theta}(\tau,t)$, with $\langle \mathbf{f}_p(t) \rangle = \mathbf{0}$ and $\langle f_{p\alpha}(t) f_{q\beta}(t) \rangle = 2\zeta_p k_B T \delta_{\alpha\beta} \delta_{pq} \delta(t - t')$. (The case $p = 0$ is not directly relevant to the present calculations.)

When a polymer is non-free-draining, i.e., when the motion of a monomer at $\mathbf{r}(\tau)$ perturbs the motion of a monomer at $\mathbf{r}(\tau')$, Eq. (2) must be amended to[22]

$$\frac{\partial \mathbf{r}(\tau,t)}{\partial t} = \int_0^N d\tau' \mathbf{H}(\mathbf{r}(\tau), \mathbf{r}(\tau')) \left( k_0 \frac{\partial^2 \mathbf{r}(\tau,t)}{\partial \tau^2} + \boldsymbol{\theta}(\tau,t) \right). \qquad (4)$$

Here, $\mathbf{H}(\mathbf{r}(\tau), \mathbf{r}(\tau'))$ is the hydrodynamic interaction matrix, which for *bulk* solutions is given by[22]

$$\mathbf{H} = \frac{1}{8\pi\eta_s}\left(\frac{\mathbf{1}}{r_{\tau\tau'}} + \frac{\mathbf{r}_{\tau\tau'}\mathbf{r}_{\tau\tau'}}{r_{\tau\tau'}^3}\right) \tag{5}$$

where $\eta_s$ is the solvent viscosity, $\mathbf{1}$ is the unit tensor, and $\mathbf{r}_{\tau\tau'} = \mathbf{r}(\tau) - \mathbf{r}(\tau')$, with $r_{\tau\tau'} = |\mathbf{r}_{\tau\tau'}|$. Equation (4) (with $\mathbf{H}$ given by Eq. (5)) can now no longer be decoupled into independent modes by the transformation that leads to Eq. (3). This difficulty can be overcome, however, by introducing an approximation (due originally to Zimm) in which $\mathbf{H}$ is replaced by a "pre-averaged" hydrodynamic interaction $h$, defined as[22]

$$h(\tau,\tau') \equiv \int d\mathbf{r}\int d\mathbf{r}'\,\Psi_{eq}(\mathbf{r},\mathbf{r}')\mathbf{H}(\mathbf{r},\mathbf{r}') \tag{6}$$

where $\Psi_{eq}(\mathbf{r},\mathbf{r}')$ is the equilibrium distribution of the monomer coordinates $\mathbf{r}$ and $\mathbf{r}'$. When $\mathbf{H}$ is given by Eq. (5), one can show that the pre-averaged interaction is

$$h(\tau,\tau') = \frac{1}{\eta_s l_0 (6\pi^3 |\tau-\tau'|)^{1/2}}. \tag{7}$$

The use of this expression in Eq. (4) (along with the expansions $\mathbf{r}(\tau,t) = \mathbf{X}_0 + 2\sum_{p=1}^{\infty}\mathbf{X}_p(t)\cos(p\pi\tau/N)$ and $\boldsymbol{\theta}(\tau,t) = N^{-1}\sum_{p=1}^{\infty}\mathbf{f}_p(t)\cos(p\pi\tau/N)$) now converts Eq. (4) to an equation that is identical to Eq. (3) except for the replacement of the parameter $\zeta_p$ by the parameter $\zeta_p^Z \equiv \eta_s(12\pi^3 N l_0^2 p)^{1/2}$. This new equation – the Rouse-Zimm equation – forms the basis of our treatment of polymer flow through narrow capillaries.

**B. Inclusion of steady shear flow**

In the presence of an imposed solvent velocity field $\mathbf{v}(\mathbf{r})$, the chain dynamics defined by the Rouse or Rouse-Zimm models is altered, because each segment of the chain now acquires an additional velocity $\mathbf{\Gamma} \cdot \mathbf{r}(\tau,t)$, where $\mathbf{\Gamma}$ is the velocity gradient tensor. The altered dynamics, for the Rouse-Zimm model specifically, are described (in normal mode form) by the equation[3,9,22,26]

$$\frac{\partial \mathbf{X}_p(t)}{\partial t} = -\frac{k_p}{\zeta_p^Z} \mathbf{X}_p(t) + \mathbf{\Gamma} \cdot \mathbf{X}_p(t) + \frac{1}{\zeta_p^Z} \mathbf{f}_p(t). \tag{8}$$

If $\mathbf{v}(\mathbf{r})$ corresponds to steady shear flow along the $z$ direction, the Cartesian components of $\mathbf{v}$ are $v_x = 0$, $v_y = 0$ and $v_z = \dot{\gamma} y$, $\dot{\gamma}$ being the shear rate, and the velocity gradient tensor becomes[23]

$$\mathbf{\Gamma} = \dot{\gamma} \begin{pmatrix} 0 & 0 & 0 \\ 0 & 0 & 0 \\ 0 & 1 & 0 \end{pmatrix}, \tag{9}$$

Under steady state conditions, the mean square extension $\langle \mathbf{R}^2 \rangle$ of a chain that evolves according to Eqs. (8) and (9) is easily calculated from the solution of Eq. (8) and the general relation $\langle \mathbf{R}^2 \rangle = 16 \sum_{p:\text{odd}} \sum_{\alpha=x,y,z} \langle X_{p\alpha}^2 \rangle$. The result is

$$\langle \mathbf{R}^2 \rangle = 16 \sum_{p:\text{odd}} \left[ \frac{3k_B T}{k_p} + \frac{k_B T}{2k_p} \left( \frac{\dot{\gamma} \zeta_p^Z}{k_p} \right)^2 \right] \tag{10}$$

with $k_p$ defined after Eq. (3) and $\zeta_p^Z$ defined after Eq. (7). But this result shows an unphysical dependence on $N$ in the limit $\dot{\gamma} \gg 1$, which can be ascribed to the infinite extensibility of Gaussian chains. This deficiency of the present model can, however, be corrected, as discussed below.

## C. Finite Extensibility

Although a constraint of finite extensibility can be imposed *rigorously* on continuum chain models[27] (so that they conform to the behavior of real chains, which cannot be stretched indefinitely), it is often at the cost of analytical tractability. But in previous work from this group,[7] we have shown that a finitely extensible chain model that is no more difficult to treat than the Gaussian chain itself can be formulated by replacing the Hamiltonian $H_0$ of Eq. (1) by the Hamiltonian

$$H = \frac{k}{2}\int_0^N d\tau \left(\frac{\partial \mathbf{r}(\tau,t)}{\partial \tau}\right)^2 \tag{11}$$

where the new spring constant $k$ is defined as

$$k = k_0 \frac{1 - \langle \mathbf{R}^2 \rangle_0 / \langle \mathbf{R}^2 \rangle_m}{1 - \langle \mathbf{R}^2 \rangle / \langle \mathbf{R}^2 \rangle_m} \equiv k_0 b \tag{12}$$

Here $\langle \mathbf{R}^2 \rangle_0$ is the mean square end-to-end distance of the unperturbed chain, $\langle \mathbf{R}^2 \rangle$ is the mean square end-to-end distance of the chain under the prevailing kinematic conditions, and $\langle \mathbf{R}^2 \rangle_m$ is the maximum mean square end-to-end distance (which is not necessarily just *N*, since under steady shear flow the chain seldom, if ever, extends to its full length; $\langle \mathbf{R}^2 \rangle_m$ is therefore regarded as a quantity to be obtained from experiment.) The above definition of $k$ – motivated by FENE models[23] – ensures that (i) when there is no flow and $\langle \mathbf{R}^2 \rangle \to \langle \mathbf{R}^2 \rangle_0$, $k$ is given correctly by $k_0$, and that (ii) when the shear rate is high and $\langle \mathbf{R}^2 \rangle \to \langle \mathbf{R}^2 \rangle_m$, $k$ becomes infinitely large and prevents the chain from deforming further.

Because $k_0 = 3k_B T/l_0^2$, the relation $k = k_0 b$ is equivalent to $k = 3k_B T/l^2$, where $l$, defined as $l = l_0/\sqrt{b}$, can be interpreted as a rescaled Kuhn length. To obtain results with the model defined by Eq. (11), therefore, it is enough to substitute $l$ for $l_0$ in the results obtained with the model defined by Eq. (1).[9] This means, for instance, that the structure of Eq. (10) continues to hold for the steady-state mean square end-to-end distance of the finitely extensible chain, but now $\zeta_p^Z$ must be replaced by $\zeta_p^{FE} \equiv \eta_s (12\pi^3 N l_0^2 p/b)^{1/2}$ and $k_p$ must be replaced by $k_p^{FE} = 6k_B T p^2 \pi^2 b/N l_0^2$. The parameter $b$ being defined as $b \equiv \left(1 - \langle \mathbf{R}^2 \rangle_0 / \langle \mathbf{R}^2 \rangle_m \right) / \left(1 - \langle \mathbf{R}^2 \rangle / \langle \mathbf{R}^2 \rangle_m \right)$, the resulting equation is actually a polynomial equation in $\langle \mathbf{R}^2 \rangle$, which can be solved (assuming that $\langle \mathbf{R}^2 \rangle_0$ and $\langle \mathbf{R}^2 \rangle_m$ are known.) Wang and Chatterjee[9] used this method to study the effects of pre-averaged hydrodynamic interactions on chain extension under shear as a function of a dimensionless shear rate, and their results were found to be in satisfactory agreement with available experimental data. As discussed in Sec. III below, the present calculations extend this work further by calculating the same quantity for the case where the flow field is imposed in a narrow capillary.

**III. CHAIN DIMENSIONS UNDER SHEAR IN A NARROW CAPILLARY**

To determine the average size of a polymer that is subject to steady shear flow in a narrow cylinder of radius $L$, in which the direction of flow is along $z$, we assume that the only effect the boundary walls have on chain dynamics is to modify the hydrodynamic interaction $\mathbf{H}$ between different parts of the chain. This surface-modified interaction is found by solving an appropriate set of fluid mechanical equations for the solvent velocity field in a cylindrical geometry. The relevant calculations (formulated in terms of the cylindrical coordinates $(\rho, \theta, z)$) are sketched in Appendix A, where it is shown that $zz$ element of $\mathbf{H}$ (the only element needed for the present purposes) is given by

$$\mathbf{H}_{zz} = \frac{1}{\pi \eta_s L^2} \int_{-\infty}^{\infty} dk_z \sum_{m=-\infty}^{\infty} \sum_{n=1}^{\infty} \frac{(1 - k_z^2/k^2)}{k^2 J_{m+1}^2(\alpha_{mn})} J_m(\alpha_{mn}\rho/L) J_m(\alpha_{mn}\rho_0/L)$$

$$\times \exp(im(\theta - \theta_0)) \exp(ik_z(z - z_0)) \quad (13)$$

where $J_\nu(x)$ is the Bessel function of order $\nu$, $\alpha_{mn}$ is the $n$th zero of the Bessel function of order $m$, i.e., $J_m(\alpha_{mn}) = 0$, and $k^2 \equiv k_z^2 + (\alpha_{mn}/L)^2$. (This expression differs by a factor of 2 from the expression derived by Doi and Harden,[13] whose general approach we have followed.) To simplify later calculations, $\mathbf{H}_{zz}$ is now pre-averaged over the equilibrium distribution $\Psi_{eq}(\mathbf{r}, \mathbf{r}')$ appropriate to a cylindrical geometry (again, following the approach discussed in Ref. 13.) The result is $\int d\mathbf{r} \int d\mathbf{r}' \Psi_{eq}(\mathbf{r}, \mathbf{r}') \mathbf{H}_{zz}(\mathbf{r}, \mathbf{r}') = h^*(|\tau - \tau'|)$, where as shown in Appendix B, $h^*(|\tau - \tau'|)$, in the narrow capillary limit $L \ll 1$ and under the so-called ground state dominance approximation,[13] is given by

$$h^*(|\tau - \tau'|) = \frac{1}{\eta_s^* l_0 (6\pi^3 |\tau - \tau'|)^{1/2}} \quad (14)$$

where $\eta_s^* \equiv \eta_s \alpha_{01}^6 J_1^6(\alpha_{01})/12\pi Q^2$, $Q \equiv \int_0^{\alpha_{01}} dx\, x J_0^3(x)$, and $\alpha_{01}$ is the first zero of the Bessel function of order 0.

So the pre-averaged hydrodynamic interaction for flow through a narrow capillary has exactly the structure of the original Zimm interaction. This means that for a finitely extensible chain under steady shear flow in a narrow capillary, the mean square end-to-end $\langle R^2 \rangle$ can immediately be obtained from Eq. (10) as

$$\left\langle \mathbf{R}^2 \right\rangle = 16 \sum_{p:\text{odd}} \left[ \frac{3k_BT}{k_p^{FE}} + \frac{k_BT}{2k_p^{FE}} \left( \frac{\dot{\gamma}\zeta_p^{FE,C}}{k_p^{FE}} \right)^2 \right] \tag{15}$$

where $\zeta_p^{FE,C} \equiv \eta_s^* (12\pi^3 N l_0^2 p/b)^{1/2}$. As noted before, this equation is actually a polynomial equation in $b$ (or $\left\langle \mathbf{R}^2 \right\rangle$). By evaluating the sums in Eq. (15) over the modes $p$ using the definition of the Riemann zeta function, and then substituting the definition of $b$, viz., $b = \left(1 - \left\langle \mathbf{R}^2 \right\rangle_0 / \left\langle \mathbf{R}^2 \right\rangle_m \right) / \left(1 - \left\langle \mathbf{R}^2 \right\rangle / \left\langle \mathbf{R}^2 \right\rangle_m \right)$, this polynomial equation is found to be

$$b^4 - b^3 - 4\varepsilon\beta^2 / 9 = 0 \tag{16}$$

where $\varepsilon = \left\langle \mathbf{R}^2 \right\rangle_0 / \left\langle \mathbf{R}^2 \right\rangle_m$, $\beta = \left( \eta_s^* \dot{\gamma} N^{3/2} l_0^3 \sqrt{31\varsigma(5)/32} / \pi^{3/2} k_B T \right)$ and $\varsigma(\cdots)$ is the Riemann zeta function. Since the factor $\tau_r \equiv \eta_s N^{3/2} l_0^3 / \sqrt{3\pi} k_B T$ defines the longest relaxation time of a Rouse-Zimm chain,[22] the parameter $\beta$ can be written as $\beta = \chi \text{Wi}$, where $\chi = \alpha_{01}^6 J_1^6(\alpha_{01}) \sqrt{31\varsigma(5)/6} / 16\pi^2 Q^2$, and Wi, the Weissenberg number, is defined as $\text{Wi} = \dot{\gamma}\tau_r$.

Equation (16) can actually be solved analytically[28] (using Mathematica), but it proves to be much more convenient to simply find the relevant root numerically (again, using Mathematica.) With $b$ in hand, the fractional chain extension, $x \equiv \left\langle R^2 \right\rangle^{1/2} / N$, is calculated from

$$x = x_0 \left[ \frac{1}{b} + \frac{4\chi^2 \text{Wi}^2}{9b^4} \right]^{1/2}, \tag{17}$$

where the parameter $x_0$ is defined as $x_0 = \left\langle R^2 \right\rangle_0^{1/2} / N$. Equation (17) is the key result of our calculations.

## IV. DISCUSSION

We are presently not aware of experimental results on polymer flow through a narrow capillary that can be directly compared with the theoretical result of Eq. (17). Under the circumstances, the measurements by Fang et al.[20] on the stretching under shear of $20\,\mu m$ $\lambda-$phage DNA at different distances $H$ from a flat glass microscope slide, being among the few to quantify the variation of chain extension with the Weissenberg number and the degree of confinement, seem to represent the best set of data for the comparison, even though they are based on a plane rather cylindrical flow geometry. The comparison is shown in Fig. 1, where the open symbols correspond to experimental data points *reconstructed* from the data points of Fig. (3) of Ref. (20) using the program Windig (downloaded from the web[29]) to estimate their $x$ and $y$ coordinates. Different symbols correspond to the different heights $H$ above the surface at which measurements of chain extension versus Weissenberg number were made, squares corresponding to $H = 10\,\mu m$, inverted triangles to $H = 7\,\mu m$, circles to $H = 5\,\mu m$, triangles to $H = 3\,\mu m$, and diamonds to $H = 1\,\mu m$. The full lines are the corresponding theoretical predictions obtained from Eq. (17) after estimating $\varepsilon = \langle \mathbf{R}^2 \rangle_0 / \langle \mathbf{R}^2 \rangle_m$ and $x_0 = \langle R^2 \rangle_0^{1/2} / N$ from the experimental data of Ref. (20) for each value of $H$ (1, 3, 5, 7 and 10 $\mu m$.) For these values of $H$, our estimates of $\varepsilon$ are, respectively, 0.73, 0.78, 0.16, 0.034 and 0.029, those of $x_0$ are, respectively, 0.062, 0.090, 0.077, 0.062 and 0.062, and the corresponding curves are colored black, blue, magenta, green and red. At the smallest values of $H$ (1 and 3 $\mu m$) the agreement between experiment and theory is quite close, despite the differences in flow geometries. Not surprisingly, there are very significant deviations between the two at larger values of $H$, since the theoretical result explicitly assumes a narrow capillary limit, and cannot strictly be applied when this condition is not met.

But the theory suggests how proximity to a surface can reduce chain extension under shear by modifying the nature of the hydrodynamic interactions between different parts of the chain.

The extent of this modification can be better appreciated by considering what the model says about chain extension when hydrodynamic interactions are included but the confining surface is not (the problem considered by Wang and Chatterjee.[9]). This is easily determined, because the required expression for the chain's end-to-end distance is also Eq. (15), but with $\eta_s^*$ (see the definition after Eq. (14)) replaced by $\eta_s$. So the equation that fixes the value of the extensibility factor $b$ continues to be given by Eq. (16) but with the parameter $\chi$ (see the definition after Eq. (16)) now defined as $\sqrt{279\varsigma(5)/96\pi^2} \equiv \chi'$, That is

$$b^4 - b^3 - 4\varepsilon\chi'^2 \mathrm{Wi}^2/9 = 0 \qquad (18)$$

with $\varepsilon = \langle \mathbf{R}^2 \rangle_0 / \langle \mathbf{R}^2 \rangle_m$ as before. Similarly the equation for the fractional extension $x$ continues to be given by Eq. (17) but with $\chi$ replaced by $\chi'$. After a slight rearrangement, this equation is

$$y \equiv \frac{\langle R^2 \rangle^{1/2}}{\langle R^2 \rangle_0^{1/2}} = \left[\frac{1}{b} + \frac{4\chi'^2 \mathrm{Wi}^2}{9b^4}\right]^{1/2} \qquad (19)$$

Figure 2 shows a graph of $y$ versus Wi (the magenta colored curve) as determined from Eq. (19) after setting $\varepsilon$ to the arbitrary value of 0.2, and after calculating $b$ from Eq. (18). The same figure includes a graph of $y$ versus Wi as determined from Eq. (17) (the blue colored curve), after setting $\varepsilon$ to the same value, and after calculating $b$ from Eq. (16). The curves clearly demonstrate the significant impact that confinement can have on chain hydrodynamics, and they lend further support to the picture proposed in the Introduction to explain the reduction in fractional extension of chains under flow near surfaces.

It should be noted, however, that the theoretical curves in Fig. 1 are sensitive to the value of $\varepsilon$ estimated from experiment. These estimates, however, are prone to error, since some subjectivity is involved in judging where on the ordinate the experimental curves of $x$ versus Wi saturate (if they do so at all.) A further complication is the lack of a data point at $Wi = 0$ for all but the $H = 1 \, \mu m$ experiment, which affects our estimates of $x_0$. These caveats aside, we believe the present formalism represents a fruitful analytical approach to the study of confined polymer dynamics that is simple, rigorous and general, and that we expect will be applicable to other solvent velocity profiles and other confinement geometries.

**APPENDIX A. CALCULATION OF THE HYDRODYNAMIC MATRIX**

When a force $\mathbf{F}$ is applied to a fluid element at $\mathbf{r}_0$, the velocity $\mathbf{v}$ imparted to the fluid element at $\mathbf{r}$ is given by

$$\mathbf{v}(\mathbf{r}) = \mathbf{H}(\mathbf{r},\mathbf{r}_0) \cdot \mathbf{F}(\mathbf{r}_0), \tag{A1}$$

which can be regarded as the defining relation of the hydrodynamic interaction $\mathbf{H}$. To find $\mathbf{H}$, the velocity $\mathbf{v}$ is assumed to satisfy both the Navier-Stokes equation in the creeping flow approximation[30]

$$\eta_s \nabla_\mathbf{r}^2 \mathbf{v} - \nabla_\mathbf{r} P = -\mathbf{F} \delta(\mathbf{r} - \mathbf{r}_0) \tag{A2a}$$

as well as the equation

$$\nabla_\mathbf{r} \cdot \mathbf{v} = 0 \tag{A2b}$$

expressing the incompressibility of the fluid. In Eq. (A2a), $\eta_s$ is the viscosity of the solvent and $P$ is its pressure field. These equations must be solved subject to the requirement that $\mathbf{v}$ vanish at the radius $L$ of the cylinder through which the fluid flows. The process begins by taking the divergence of Eq. (A2a) and using Eq. (A2b) to eliminate **v**. The resulting equation for $P$ is

$$\nabla_r^2 P = \nabla_r \cdot \mathbf{F}\delta(\mathbf{r}-\mathbf{r}_0), \tag{A3}$$

which is solved by

$$P = \int d\mathbf{r}' G(\mathbf{r},\mathbf{r}')\nabla_{r'} \cdot \mathbf{F}\delta(\mathbf{r}'-\mathbf{r}_0) \tag{A4}$$

where the Green's function $G$ is the solution of

$$\nabla_r^2 G(\mathbf{r},\mathbf{r}_0) = \delta(\mathbf{r}-\mathbf{r}_0) \tag{A5}$$

satisfying the same boundary conditions as **v**. Using cylindrical coordinates $\mathbf{r}=(\rho,\theta,z)$, this solution is found to be

$$G(\mathbf{r},\mathbf{r}_0) = -\frac{1}{\pi}\int_{-\infty}^{\infty} dk_z \sum_{m,n} \frac{1}{k^2 L^2 J_{m+1}^2(\alpha_{mn})} J_m(\alpha_{mn}\rho/L) J_m(\alpha_{mn}\rho_0/L) e^{im(\theta-\theta_0)} e^{ik_z(z-z_0)} \tag{A6}$$

where $k^2 = k_z^2 + (\alpha_{mn}/L)^2$, and $\alpha_{mn}$ is the $n$th zero of the Bessel function of order $m$. The substitution of (A6) in (A4) determines the pressure field as

$$P = -\frac{1}{\pi}\int_{-\infty}^{\infty} dk_z \sum_{m,n} \frac{J_m(\alpha_{mn}\rho/L)}{k^2 L^2 J_{m+1}^2(\alpha_{mn})} e^{im(\theta-\theta_0)} e^{ik_z(z-z_0)} \times$$

$$\times\left[-\frac{\alpha_{mn}}{L} F_\rho J_m'(x_0) + \frac{im}{\rho_0} F_\theta J_m(x_0) + ik_z F_z J_m(x_0)\right] \tag{A7}$$

where $x_0 = \alpha_{mn}\rho_0/L$, the prime on the Bessel function denotes differentiation with respect to its argument, and $F_\rho$, $F_\theta$ and $F_z$ are the components of **F** along the cylindrical axes.

Since flow has been assumed to occur along the $z$ direction, it is the $z$ component of the fluid velocity, $v_z$, and its relation to the $z$ component of the force, $F_z$, that is needed in the determination $\mathbf{H}_{zz}$. To this end, the component $v_z$ is expressed as an eigenfunction expansion:

$$v_z = \frac{1}{2\pi} \int_{-\infty}^{\infty} dk_z \sum_{m,n} C_{mn} J_m(\alpha_{mn}\rho/L) \, e^{im(\theta-\theta_0)} e^{ik_z(z-z_0)} \tag{A8}$$

where the $C_{mn}$ is an unknown expansion coefficient. A similar eigenfunction expansion exists for the delta function, which from the equation for the Green's function [Eq. (A5)] can be shown to be

$$\frac{1}{\rho}\delta(\rho-\rho_0)\delta(\theta-\theta_0)\delta(z-z_0) = \frac{1}{\pi L^2} \int_{-\infty}^{\infty} dk_z \sum_{m,n} \frac{1}{J_{m+1}^2(\alpha_{mn})} J_m(\alpha_{mn}\rho/L) J_m(\alpha_{mn}\rho/L) \times$$

$$\times e^{im(\theta-\theta_0)} e^{ik_z(z-z_0)} \tag{A9}$$

By substituting Eqs. (A7), (A8) and (A9) into Eq. (A2a), the coefficient $C_{mn}$, and hence $v_z$ itself, can be determined, and this in turn identifies $\mathbf{H}_{zz}$ as the expression given in Eq. (13).

**APPENDIX B. THE PREAVERAGED HYDRODYNAMIC INTERACTION**

The pre-averaging of $\mathbf{H}_{zz}$ requires an expression for the equilibrium pair distribution function $\Psi_{eq}(\mathbf{r},\mathbf{r}')$. This function separates into transverse and longitudinal components. That is,

$$\Psi_{eq}(\mathbf{r},\mathbf{r}') = \Psi_{\parallel}(z,z'|\tau,\tau')\Psi_{\perp}(\rho,\theta;\rho',\theta'|\tau,\tau') \tag{B1a}$$

where

$$\Psi_{\parallel}(z,z'|\tau,\tau') = \sqrt{\frac{3}{2\pi d_0^2}} \exp\left(-\frac{3(z-z')^2}{2l_0^2|\tau-\tau'|}\right) \tag{B1b}$$

and

$$\Psi_{\perp}(\rho,\theta;\rho',\theta'|\tau,\tau') = C^{-1} \int_0^L d\rho_N \rho_N \int_0^{2\pi} d\theta_N \int_0^L d\rho_0 \rho_0 \int_0^{2\pi} d\theta_0 G_0(\rho_N,\theta_N;\rho,\theta|N,\tau) \times$$

$$\times G_0(\rho,\theta;\rho',\theta'\,|\,\tau,\tau')G_0(\rho',\theta';\rho_0,\theta_0\,|\,\tau') \tag{B1c}$$

where $C = \int_0^L d\rho_N \rho_N \int_0^{2\pi} d\theta_N \int_0^L d\rho_0 \rho_0 \int_0^{2\pi} d\theta_N G_0(\rho_N,\theta_N;\rho_0,\theta_0\,|\,N)$ is a normalization factor, and $G_0$ is the solution of

$$\left(\frac{\partial}{\partial N} - \frac{l_0^2}{6}\nabla_\perp^2\right)G_0(\rho,\theta;\rho',\theta'\,|\,N) = \frac{1}{\rho}\delta(\rho-\rho')\delta(\rho-\rho')\delta(N) \tag{B1d}$$

satisfying the boundary condition that $G_0$ vanish at the walls of the cylinder. The solution is

$$G_0(\rho,\theta;\rho',\theta'\,|\,N) = \frac{1}{\pi L^2}\sum_{m,n}\frac{1}{J_{m+1}^2(\alpha_{mn})}J_m(\alpha_{mn}\rho/L)J_m(\alpha_{mn}\rho'/L)e^{im(\theta-\theta')} \times$$

$$\times e^{-\alpha_{mn}^2 N l_0^2 / 6L^2} \tag{B2}$$

After using this expression in Eq. (B1c), along with the general relations $\int_0^{2\pi}d\theta\,e^{im\theta} = 2\pi\delta_{m,0}$ and $\int_0^L d\rho\,\rho\,J_0(\alpha_{mn}\rho/L) = (L^2/\alpha_{mn})J_1(\alpha_{mn})$, and then invoking the assumption of ground state dominance, in which only the leading order term in the expansion of $G_0$ is retained, one finds, after routine algebra, that

$$\Psi_\perp(\rho,\theta;\rho',\theta'\,|\,\tau,\tau') = \frac{J_0^2(\alpha_{01}\rho/L)J_0^2(\alpha_{01}\rho'/L)}{\pi^2 L^4 J_1^4(\alpha_{01})} \tag{B3}$$

independent of $\tau$ and $\tau'$. Equations (B3) and (B1b), along with Eq. (13), are now substituted into the definition of the pre-averaged hydrodynamic interaction. Invoking the ground state dominance approximation again,[13] and carrying out the simple integral over $(z-z')$, one obtains

$$\langle \mathbf{H}_{zz}\rangle = \frac{4}{\eta_s \pi L^2 \alpha_{01}^4 J_1^6(\alpha_{01})}Q^2 K \tag{B4}$$

where $Q$ is the integral defined after Eq. (14) and $K$ is the integral

$$K = \int_{-\infty}^{\infty} dk_z \frac{1}{k^2}\left(1 - \frac{k_z^2}{k^2}\right)\exp\left(-k_z^2 l_0^2 |\tau - \tau'|/6\right) \tag{B6}$$

with $k^2 = k_z^2 + (\alpha_{01}/L)^2$. This integral can be evaluated with Mathematica; the result is

$$K = \frac{\sqrt{\pi}}{2B}\left[2AB - \sqrt{\pi}\left(2A^2B^2 - 1\right)e^{A^2B^2}\operatorname{erfc}(AB)\right] \tag{B7}$$

where $A = l_0\sqrt{|\tau - \tau'|/6}$ and $B = \alpha_{01}/L$. In the narrow capillary limit $L \ll 1$, the error function in this expression can be approximated by its asymptotic expansion,[31] whereupon

$$K \approx \frac{\sqrt{6\pi}}{2l_0(\alpha_{01}/L)^2}\frac{1}{|\tau - \tau'|^{1/2}} \tag{B8}$$

After substituting Eq. (B8) into Eq. (B4) and rearranging terms, one obtains Eq. (14).

**FIGURE CAPTIONS**

1. Experimental and theoretical curves of the mean fractional extension $x$ as a function of the Weissenberg number Wi. The open symbols correspond to experimental data points reconstructed from the data points in Fig. 3 of Ref. (20), which were obtained from measurements on the extension of $20\,\mu$m $\lambda-$phage DNA under shear at different distances $H$ from a glass surface. Different symbols correspond to the different values of $H$, squares corresponding to $H=10\,\mu$m, inverted triangles to $H=7\,\mu$m, circles to $H=5\,\mu$m, triangles to $H=3\,\mu$m, and diamonds to $H=1\,\mu$m. The full lines are the corresponding theoretical predictions obtained from Eq. (17) after estimating $\varepsilon = \langle \mathbf{R}^2 \rangle_0 / \langle \mathbf{R}^2 \rangle_m$ and $x_0 = \langle R^2 \rangle_0^{1/2} / N$ from the experimental data of Ref. (20) for each value of $H$ (1, 3, 5, 7 and 10 $\mu$m.) For these values of $H$, the estimates of $\varepsilon$ are, respectively, 0.73, 0.78, 0.16, 0.034 and 0.029, those of $x_0$ are, respectively, 0.062, 0.090, 0.077, 0.062 and 0.062, and the corresponding curves are colored black, blue, magenta, green and red.

2. Extension ratio $y \equiv x/x_0$ as a function of Weissenberg number, Wi, for confined and unconfined polymers, as calculated, respectively, from Eqs. (17) and (19). at an arbitrary value of $\varepsilon = 0.2$. The blue curve corresponds to the confined polymer and the magenta curve to the unconfined polymer.

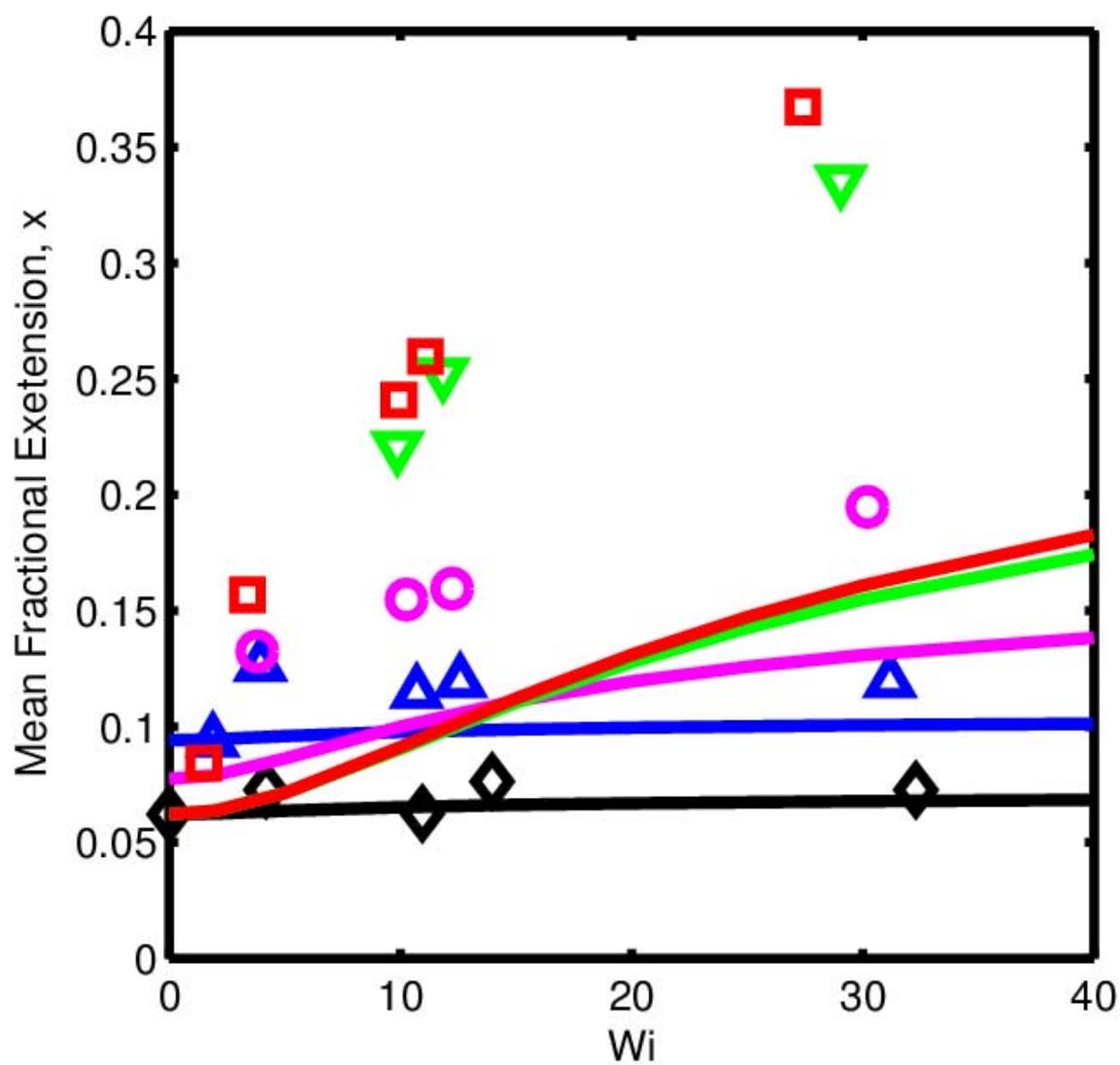

**FIGURE 1**

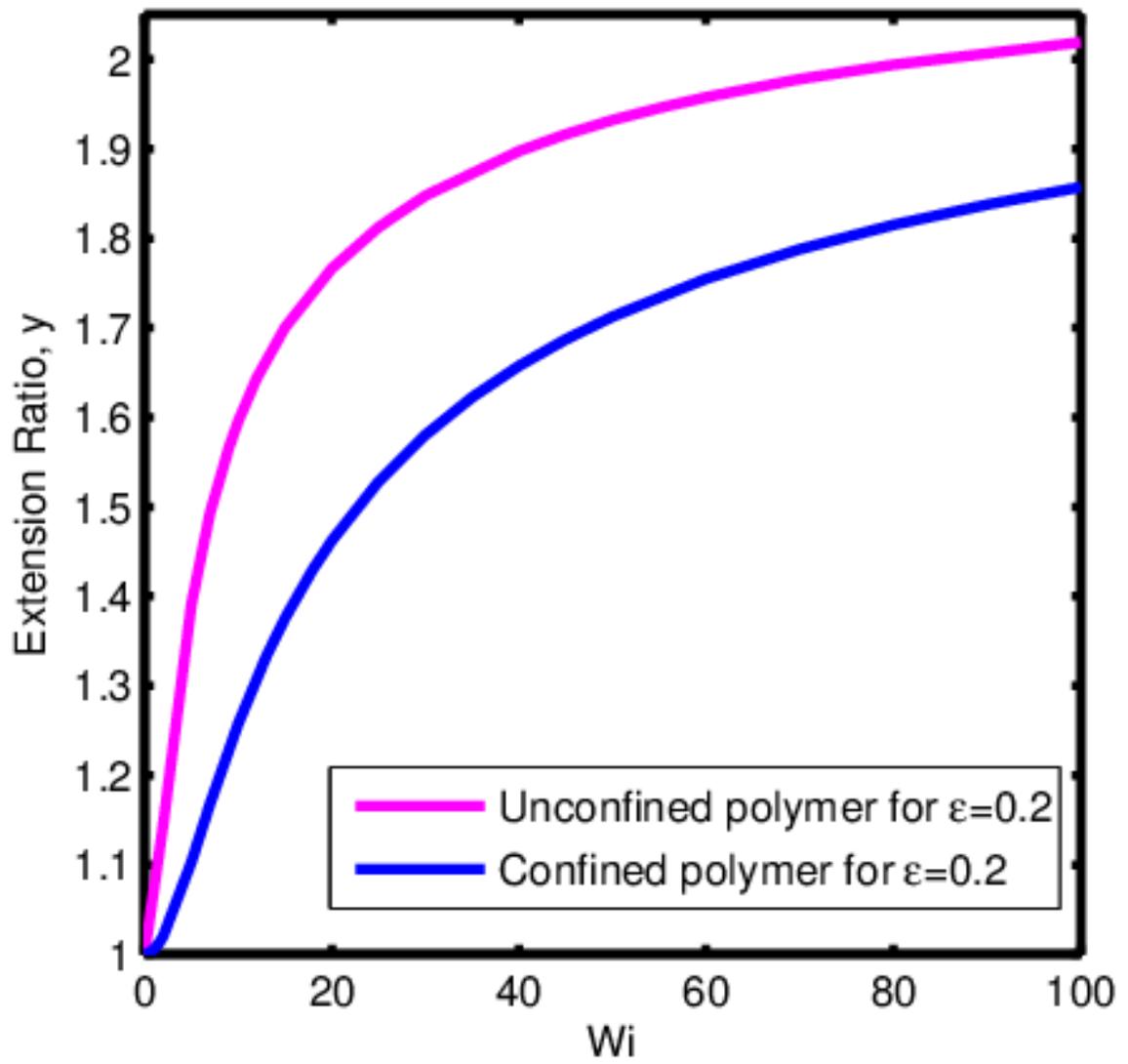

**FIGURE 2**